\documentclass[a4paper,10pt, onecolumn]{article}
\usepackage[utf8]{inputenc}
\usepackage{amsfonts}
\usepackage{amsmath}
\usepackage{amssymb}
\usepackage{graphicx}
\usepackage{dcolumn}
\usepackage{bm}
\usepackage{longtable}
\usepackage[usenames, dvipsnames, table]{xcolor}
\usepackage{authblk}
\usepackage{geometry} 
\title{Mesoscopic Community Structure of Financial Markets Revealed by Price and Sign Fluctuations}

\author[1,\footnote{Correspondence to almog@lorentz.leidenuniv.nl}]{A. Almog}
\author[1]{F. Besamusca}
\author[1]{M. MacMahon}
\author[1]{D. Garlaschelli}
\affil[1]{Instituut-Lorentz for Theoretical Physics, Leiden Institute of Physics, \newline University of Leiden, Niels Bohrweg 2, 2333 CA Leiden, The Netherlands}

\date{}

\begin{document}

    \maketitle
    
\begin{abstract}
The mesoscopic organization of complex systems, from financial markets to the brain, is an intermediate between the microscopic dynamics of individual units (stocks or neurons, in the mentioned cases), and the macroscopic dynamics of the system as a whole. The organization is determined by ``communities'' of units whose dynamics, represented by time series of activity, is more strongly correlated internally than with the rest of the system. Recent studies have shown that the binary projections of various financial and neural time series exhibit nontrivial dynamical features that resemble those of the original data. This implies that a significant piece of information is encoded into the binary projection (i.e. the sign) of such increments.
Here, we explore whether the binary signatures of multiple time series can replicate the same complex community organization of the financial market, as the original weighted time series. 
We adopt a method that has been specifically designed to detect communities from cross-correlation matrices of time series data. Our analysis shows that the simpler binary representation leads to a community structure that is almost identical with that obtained using the full weighted representation. These results confirm that binary projections of financial time series contain significant structural information. 
\end{abstract}

\section*{Introduction}

One of the most important properties of complex systems is community structure. 
Real-world complex systems are organized in a modular way, with clusters of units sharing similar dynamics or functionality.
However, while the clusters are internally cohesive, they can maintain contrasting dynamics. 
The problem of resolving and identifying these mesoscopic structures, without any prior information, is extremely challenging. 
In financial markets, the mesoscopic scale corresponds to sets of stocks that share similar price dynamics.
The knowledge of the market structure is highly valuable, and can assist in hedging risks and for better understanding of the market.   
Consequently, over the past years, scientists have deployed and developed many time series techniques to retrieve qualitative information regarding the hierarchy and structure of financial markets \cite{Sinha,Bouchaud,Mantegna1,Mantegna2}.\\

A promising approach is that of employing community detection techniques, developed in network theory \cite{Fortunato,Newman}, on empirical correlation matrices (resulted from multiple time series).
However, the methods are originally constructed to detect dense clusters of nodes within graphs (networks), and are not adapted to deal with correlation matrices. 
Recently, a novel method was proposed, which has been specifically designed to detect communities from correlation matrices of multiple time series \cite{MacMahon}. 
When applied to financial time series, the method was able to capture the dynamical modularity of real markets. 
Remarkably, the method identified clusters of stocks which are correlated internally, but are anti-correlated with each other.\\

Traditionally, the main object of time series analysis is the characterization of patterns in the amplitude of the increments of the quantities of interest (stock price in our case).
The analysis requires a weighted description of the system, i.e. both the amplitude and the sign of the activity.
Indeed, a time series of increments enclose complete information about the amplitude of the fluctuations of the original signal. 
However, a significant part of this information is encoded in the purely `binary' projection of the time series, i.e. its sign. 
Recent studies have shown various forms of statistical dependency between the sign and the absolute value of fluctuations \cite{Spada,Alexander,Boguna}.
Recently, a study has shown a robust empirical relationship between binary and non-binary properties of real financial time series \cite{Almog}. The research shows that binary signatures, which retain only the sign of fluctuations, encode significant information regarding the full behaviour of the stock (both amplitude and direction).
Motivated by these recent results, here we further explore the higher-order relations between financial time series and their corresponding binary signatures, in a more complex setting.
Here we study whether the binary signatures of assets can reproduce the same complex community organization of financial markets, as the weighted information.\\

\begin{figure}[t]
\centering
\includegraphics[width=.9\textwidth]{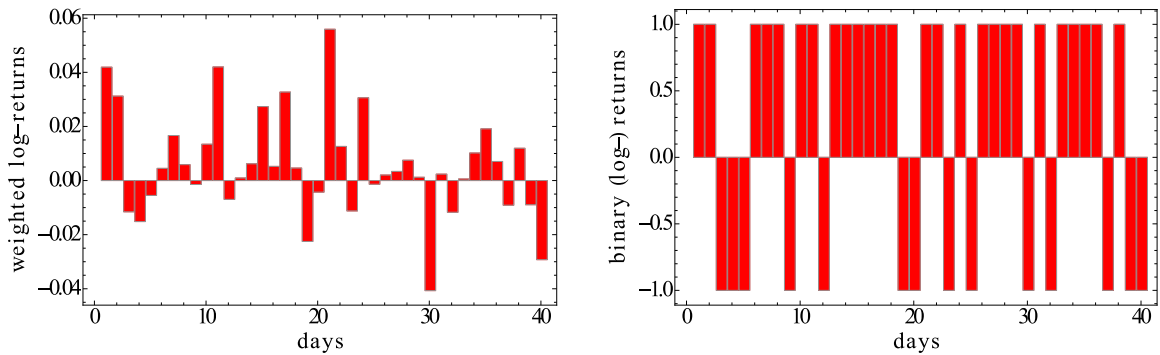}
\caption{‘Weighted’ (left) versus ‘Binary’ (right) time series of log-returns of the Apple stock over a period of 40 days starting from 7/5/2011}\label{fig:binarize}
\end{figure}

To this end, we use the daily closing prices of the stocks of three indexes (S$\&$P500, FTSE100 and NIKKEI225) over the period 2001-2011.
For each index, we restrict our sample to the maximal group of stocks that are traded continuously throughout the selected period.
This results in 445  stocks for the S$\&$P500, 78 stocks for the  FTSE100 and 193  stocks for the NIKKEI225.  
Given a stock price $P_i(t)$  where $i$ denotes one of the $N$ stocks in the index, and $t$ denotes one of the $T$ observed temporal snapshots (days), the log-return is defined as
\begin{equation}
 r_i(t)\equiv \log\left[ \frac{P_i(t+1)}{P_i(t)}\right].
\end{equation}
For each stock in the system we use the time series of it's log-returns for our analysis. This is the construct we refer to as the ``weighted time series'' throughout the rest of the paper.
In contrast, the ``binary signatures'' only reveal the direction of the fluctuation (sign) in the price and are defined as
\begin{equation}
x_i(t)\equiv \textrm{sign} [r_i(t)]=
\left\{\begin{array}{rr}
+1&r_i(t)>0\\
0&r_i(t)=0\\
-1&r_i(t)<0
\end{array}\right ..
\label{eq:sign}
\end{equation}
In fig. \ref{fig:binarize} we show a simple example of a weighted time series, along with the corresponding binary projection.\\

The two types of information are in fact different descriptions of the same system, and are used to construct cross-correlation matrices. In turn, we deploy three popular community-detection algorithms \cite{Potts1,Potts2,Louvain,Spectral} specifically adapted, where necessary, for the correct use of cross-correlation matrices \cite{MacMahon}.   
We examine and quantify similarities and variations in the organization of the markets for these two representations. 
This approach reveals some interesting results.
First, we can quantify the level of information encoded within the binary signatures, with respect to the full weighted time series. Secondly, we observe that both the binary and weighted representations yield very similar structures, which indicates that most of the information regarding the structure of financial communities is already encoded within the sign of a stock.\\

\section*{Results}

\subsection*{Spectral Analysis}
\label{SA}
\begin{figure}[t]
\centering
\includegraphics[width=.9\textwidth]{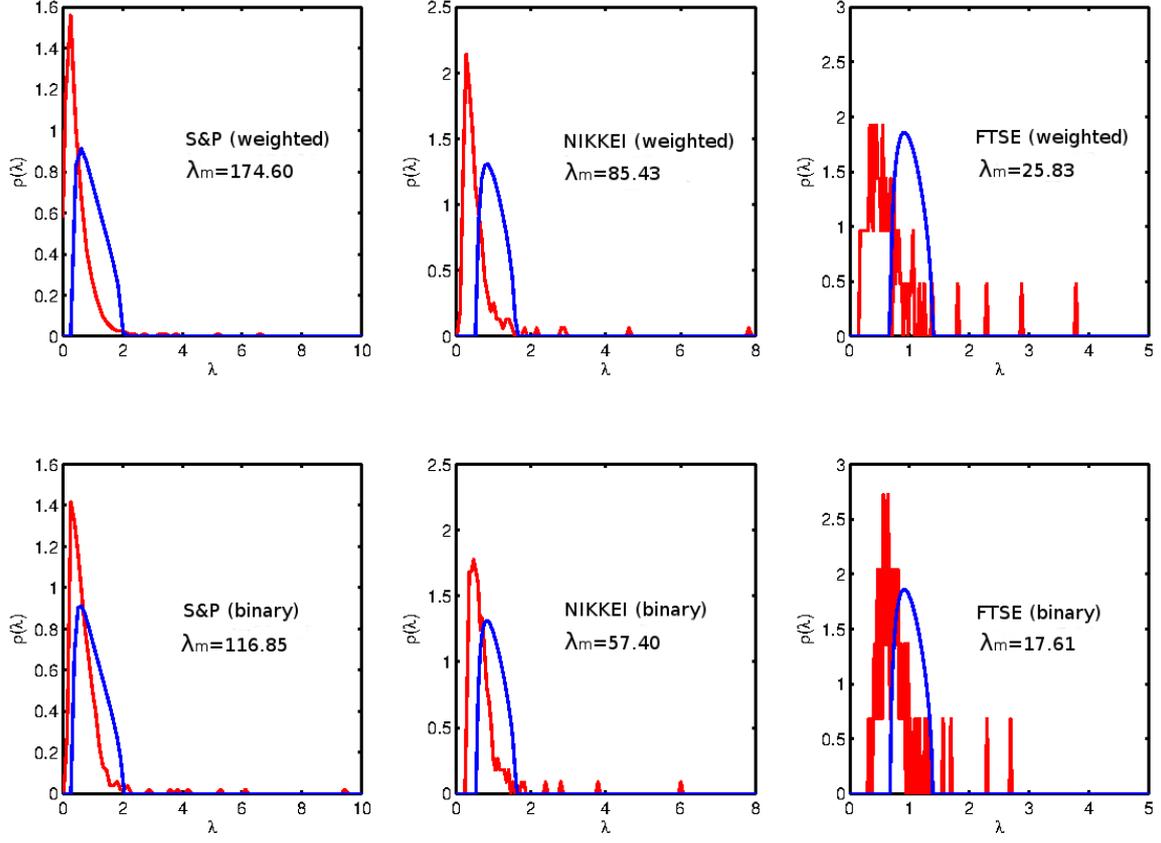}
\caption{The eigenvalue density distribution (of the cross-correlation matrix) for the different indexes, where the upper panels are for the weighted series and the lower panels are for the binary series. The red curve is the empirical eigenvalue distribution and the blue curve the Marchenko-Pastur distribution. The largest empirical eigenvalue $\lambda_m$ is not shown in the plots, but the its value is reported in each panel.}\label{Dist1}
\end{figure}
In this section we analyse the eigenvalue density distribution of the cross-correlation matrices for the two representations of the data (binary and weighted).
When plotting the density distribution one can identify specific spectral properties that have structural implications. 
In other words, it is possible to identify distinct eigenvalues in the spectrum, which correspond to correlated clusters of stocks, and typically indicate a non-trivial structure.\\  

To begin, we first need to discuss a filtering technique, based on Random Matrix Theory (RMT) \cite{Plerou,Wigner,Mehta}, which is used to identify non-random properties of empirical correlation matrices.    
The majority of the eigenvalues present in the spectrum of an empirical correlation matrix result from randomly induced correlations between the time series. 
In the generic case, where one measures the correlation between $N$ independent random time series for $T$ time steps (the observed period), then the resulting correlation matrix would be an $N$ by $N$ Wishart matrix, whose statistical properties are well known \cite{Plerou2,Laloux}. In the limits where $N,T\rightarrow \infty$ and  $T/N \geq1$ the eigenvalues of the Wishart matrix are distributed according to a Marchenko-Pastur distribution
\begin{equation}
 P(\lambda)=\frac{T}{N}\frac{\sqrt{(\lambda_{+}-\lambda)(\lambda-\lambda_{-})}}{ 2\pi \lambda }~~~~~\textrm{if}~~~~~\lambda_-\leq\lambda\leq \lambda_+
\end{equation}
and $P(\lambda)=0$ otherwise.
The boundaries $\lambda_{+}$ and $\lambda_{-}$ are dependent on the data size and given by \begin{equation}                                                                                                                                                                                        
\lambda_{\pm}=\left[1\pm\sqrt{\frac{N}{T}}\right]^2.
 \end{equation}
This analytic curve represents the boundaries of the bulk eigenvalues, which predominantly represent noise, and so have little meaning assigned to them. The eigenvalues outside this range however have structural implications, and correspond to groups of correlated stocks \cite{Laloux}.
As a result, any empirical correlation matrix $C$ can be identified as a sum of of two matrices:\\
\begin{equation}
 C=C^{(r)}+C^{(s)},
\end{equation}
where $C^{(r)}$ is the random part aggregated from the eigenvalues in the random spectrum ($\lambda_-\leq\lambda\leq \lambda_+$)
\begin{equation}
C^{(r)}\equiv \sum_{i:\lambda_-\leq\lambda_i\leq \lambda_+} \lambda_i |v_{i}\rangle \langle v_i|.
\end{equation}
We refer to $C^{(s)}$ as the ``structured'' component, which is composed from those eigenvalues above the boundary of the bulk eigenvalues. $\lambda > \lambda_+$.\\

Moving forward, in financial markets it well established that stocks typically move up or down together, an effect known as the market mode. This effect is indicated by the presence of a very large eigenvalue $\lambda_m$, orders of magnitude larger than the rest. Since this eigenvalue represents a common factor influencing all the stocks in a given market, from a structural perspective, the market mode eigenvalue signifies the presence of one single super-community, containing all the stocks in the market.\\

Thus, the other eigenvalues (not including the market mode), which deviate from the bulk, $\lambda_+<\lambda_i<\lambda_m$ are the ones corresponding to mesoscopic clusters, i.e. groups of stocks with similar dynamics. This observation results in a further decomposition of the empirical correlation matrix
\begin{equation}
 C=C^{(r)}+C^{(g)}+C^{(m)},
 \label{superposition}
\end{equation}
where
\begin{equation}
C^{(m)}\equiv \lambda_m |v_{m}\rangle \langle v_m|
\end{equation}
represents the market mode, and 
\begin{equation}
C^{(g)}\equiv \sum_{i:\lambda_{+}<\lambda_i<\lambda_{m}} \lambda_i |v_{i}\rangle \langle v_i|
\end{equation}
represents the remaining correlated groups.
These sub-groups of correlated stocks comprise the mesoscopic structure of the market. They are also referred as ``group modes'' in some of the literature \cite{Sinha,Laloux}.\\

Our focus here is to detect these eigenvalues in the spectrum of both the binary and weighted data. Moreover, we want to explore the similarities and differences between the two spectra to inform us about the corresponding structures yielded by each type of data.\\ 
\begin{figure}[t]
\centering
\includegraphics[width=.9\textwidth]{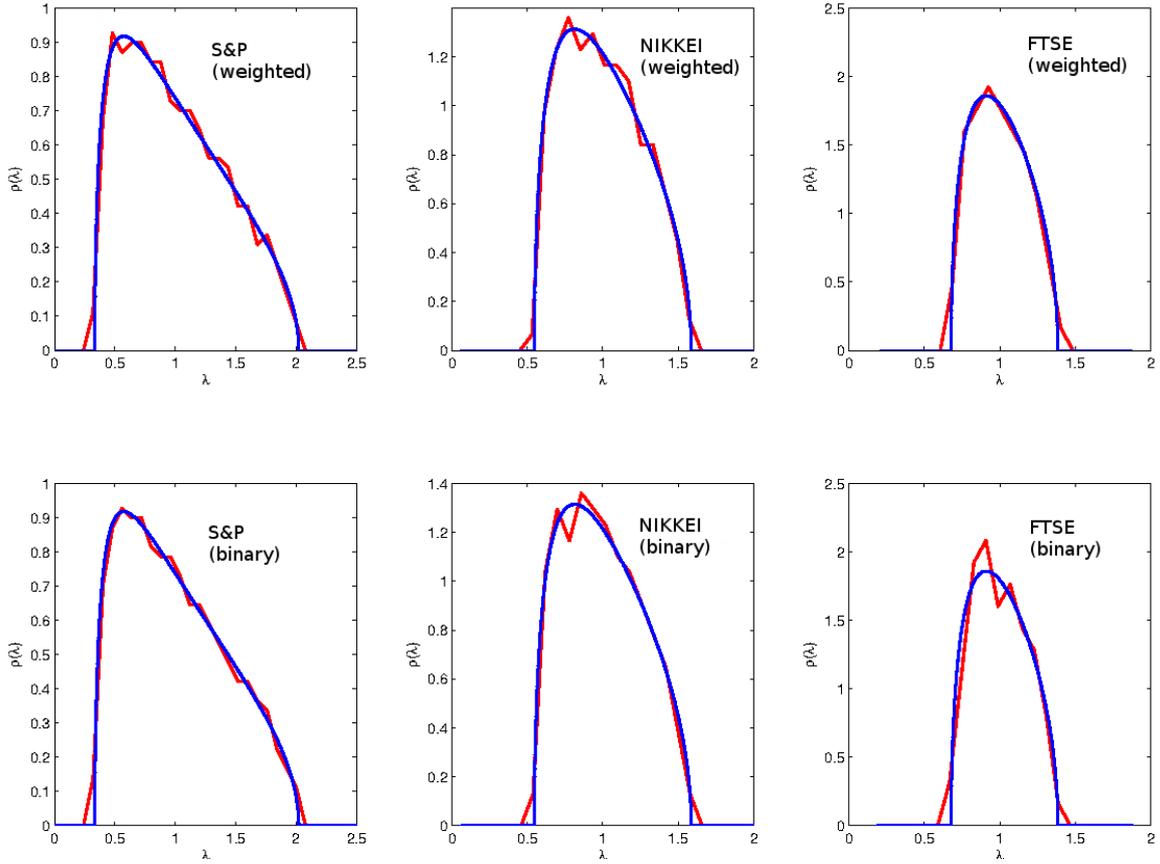}
\caption{The eigenvalue density distribution of the Pearson correlation matrices where the upper panels are for the weighted series and the lower panels are for the binary series. The red curve is the empirical eigenvalue distribution and the blue curve the Marchenko-Pastur distribution.}\label{Dist2}
\end{figure}

In Figure \ref{Dist1} we plot the eigenvalue density distribution for the three different indices.
The top row corresponds to the weighted representation (log-returns), and the bottom row corresponds to the binary representation (binary signatures). We can observe the known structure of the financial markets in the weighted data, however this complex structure also exists in the binary data. This result is non trivial. We can observe a market mode, and several deviating eigenvalues also in the ``simpler'' binary data (with the same order of magnitude).\\ 

We also want to inspect whether both descriptions of the system function the same under randomization.
The returns of each stock were separately permuted randomly, therefore preserving the total return of the stocks and destroying the daily correlation between the returns. Once the time series entries are shuffled, both binary and weighted correlation matrices end up as elementary random matrices.
As discussed before, the eigenvalues of such matrices will be distributed with a  Marchenko-Pastur distribution.

In Figure \ref{Dist2} we plot the density distribution of the shuffled data for the three different indices. 
The top row corresponds to the weighted representation (log-returns), and the bottom row corresponds to the binary representation (binary signatures).
As expected, in both cases we observed the known characteristics of a random matrix.
The spectra of both representations collapsed to the known analytic curve.\\

To sum up this section, we identified a sub-group structure both in the weighted and the binary representation of the three indices.
Each of the binary spectra we studied retain all the known properties of a ``regular'' (weighted) spectrum (random bulk, market mode and group modes). This result propels us to do a more refined analysis, and to further explore (and compare) the sub-group structure of the different indices. In the next section we will apply community detection algorithms to extract a more detailed structure for both representations, so that we can better quantify the similarities and the variations.

\subsection*{Community Structure}


\begin{table}[t]
\center
\begin{tabular}{|lc|lc|}
\hline
Consumer Discretionary: &\color{Purple} $\blacksquare$ \color{black} & Consumer Staples: &\color{Aquamarine}$\blacksquare$ \color{black}\\
Energy: &\color{CadetBlue} $\blacksquare$ \color{black} & Financials: &\color{green} $\blacksquare$ \color{black}\\
Health Care: &\color{Red} $\blacksquare$ \color{black} & Industrials: &\color{Orange} $\blacksquare$ \color{black}\\
Information Technology: &\color{Blue} $\blacksquare$ \color{black} & Materials: &\color{Yellow} $\blacksquare$ \color{black}\\
Telecom. Services: &\color{Magenta} $\blacksquare$ \color{black} & Utilities: &\color{Brown} $\blacksquare$ \color{black}\\
\hline
\end{tabular}
\caption{The 10 industry sectors in the Global Industry Classification Standard (GICS), with the color representation used to highlight the sectors in the following Figures.}
\label{tbl:GICScolors}
\end{table}

\begin{figure}[t]
\centering
\includegraphics[width=.9\textwidth]{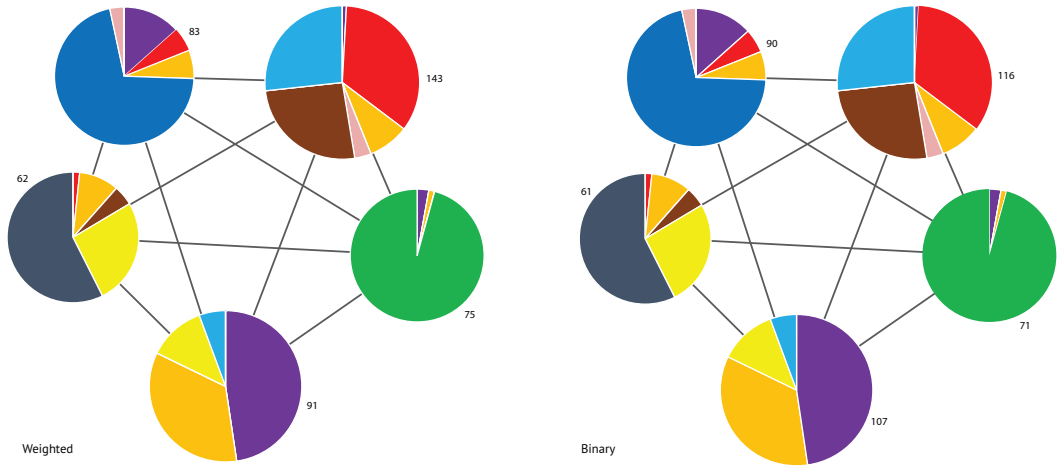}
\caption{Communities of the S\&P 500 (daily closing prices from 2001 to 2011) generated using the modified Louvain algorithm \cite{MacMahon}. Each community is labelled with the number of stocks and the pie chart represents the relative composition of each community based on the industry sectors of the constituent stocks (color legend in Table I). The inter-community link weights are negative, indicating that
the communities are all residually anti-correlated.}
\label{SP-CD}
\end{figure}

In network theory, a community structure is the partition of the network into relatively dense sets of nodes, with respect to the rest of the network. More specifically, it is the organization into clusters of nodes with dense connections internally, while the connections between the clusters are sparser. Community detection is the identification of such clusters of agents (nodes) in the system (network). In the last decade there has been a burst of research concerning this topic, across a myriad of different fields \cite{Fortunato}.

\begin{figure}[htb]
\centering
\includegraphics[width=.9\textwidth]{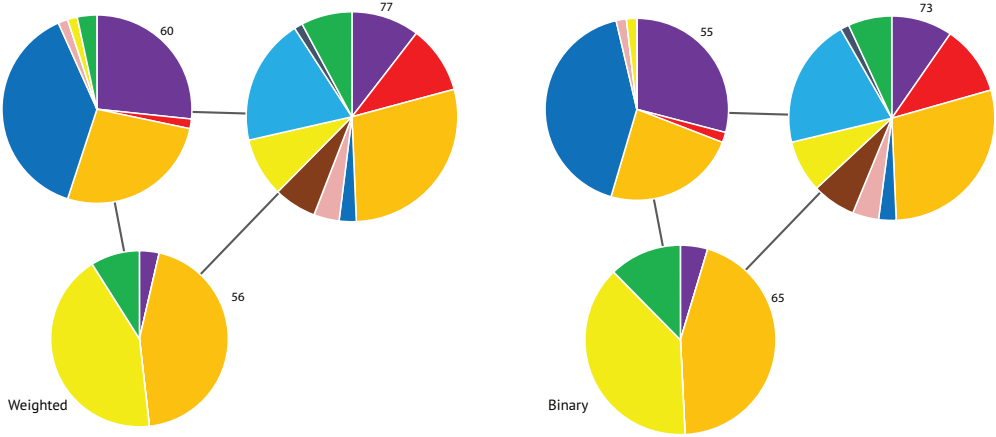}
\caption{Communities of the Nikkei 225 (daily closing prices from 2001 to 2011) generated using
the modified Louvain algorithm \cite{MacMahon}. Each community is labelled with the number of stocks, and the pie chart represents the relative composition of each community based on the industry sectors of the constituent stocks (colour legend in Table I).The link weights are negative, indicating that the communities are all residually anti-correlated. }
\label{NIKKEI-CD}
\end{figure}

This promising approach has also been applied to analyse time series data \cite{saramaki_correlations,mason_correlations,isogai,Piccardi} where the goal is to identify clusters of components with a similar dynamics.
The attempts made so far have basically replaced network data with cross-correlation matrices as the input.
However, this procedure suffers from a significant limitation.
The null hypotheses used in the network-based algorithms are inconsistent with the properties of correlation matrices. As a result, these approaches can introduce an undesired bias when applied to the detection of communities in time series based networks.\cite{MacMahon}.
Here we adopt a new method \cite{MacMahon}, which is specifically shaped to deal with correlation matrices, based on the spectral properties we presented in the previous section. 
The method presents an improved and consistent way to cluster multiple time series, by leveraging a set of null models, specifically designed for use with correlation matrices.\\

Applying the new approach, we use three popular community detection algorithms, customizing where necessary to be effective with correlation matrices. The three algorithms we use in this paper are known as the Potts (or spin glass) method \cite{Potts1,Potts2}, the Louvain method \cite{Louvain} and the spectral method \cite{Spectral}, and are modified in \cite{MacMahon} to correctly deal with correlation matrices. 
The algorithms use a modularity optimization process, where modularity is a measure for how ``optimal'' your partition is.
The algorithms attempt to choose a specific partitioning of the network into groups such that the corresponding value of the modularity is maximized (see Methods). 
The modularity implements a new null hypothesis, which is fitted to time series (correlation matrices). 
More specifically, the hypothesis considers the empirical correlation matrix as a superposition of modes (equation \ref{superposition}), and decomposes it accordingly. 
Both the random mode and market mode are filtered out, and we are left with only the informative group mode, which is then used to extract the market structure.\\ 

Thus, we end up with three community detection algorithms that are consistent with time series data and represent the counterparts of the most popular techniques used in network analysis. 
The method allows us to explore the mesoscopic structure of different financial indices, and more specifically compare the different community structures resulting from the different representations (binary and weighted).\\

First, we perform community detection on both the binary and the weighted time series, using all three community detection methods, for the full time period of the data (2001-2011). We pick the division that maximizes the modularity (for a specific representation and algorithm), and compare the results for the two types of information. In the case where several divisions maximize the modularity (different runs result in different divisions), we take the division with the most occurrences over 1000 runs (the highest probability).  Here, we want to identify groups with similar dynamics over the ten year period, with the rationale that such a long period will reduce the noise.\\

To help further explore the communities resulting from the different passes, we label each of the stocks according to its industry sector from the Global Industry Classification Standard (GICS). This classification represents a more ``traditional'' frame of mind where the different sectors are comprised of stocks conforming to a particular, qualitative description of the industry they represent. Recent results show that real markets have a more complex structure \cite{Bouchaud,Utsugi,Potters,Plerou}, where different sectors are mixed in different sub-groups, i.e. the communities are assembled out of stocks from different sectors. 
Furthermore, the classification helps us to compare the results of the two representations in a very clear and visual way.\\

\begin{figure}[htb!]
\centering
\includegraphics[width=.9\textwidth]{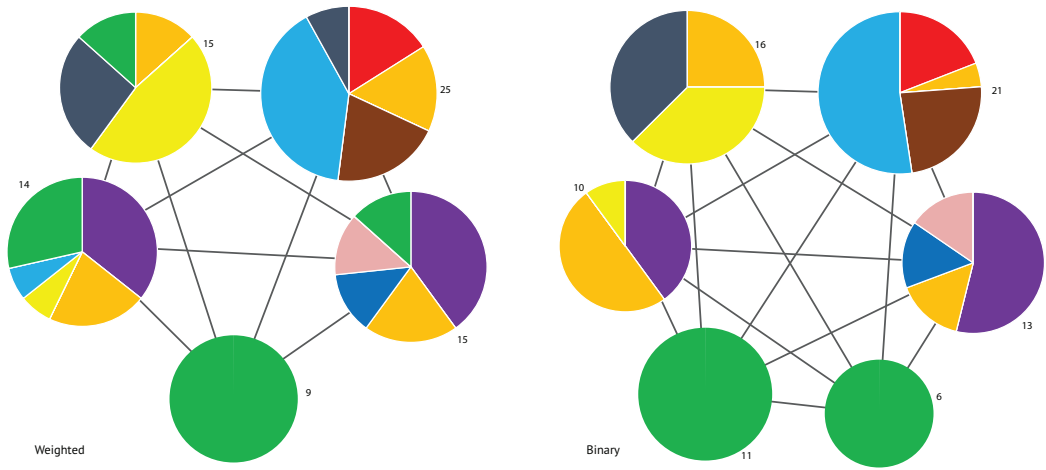}
\caption{Communities of the FTSE 100 (daily closing prices from 2001 to 2011) generated using the modified Louvain algorithm\cite{MacMahon}. Each community is labelled with the number of stocks, and the pie chart represents the relative composition of each community based on the industry sectors of the constituent stocks (color legend in Table I). The inter-community link weights are negative, indicating that the communities are all residually anti-correlated.}
\label{FTSE-CD}
\end{figure}

In Figures \ref{SP-CD} and \ref{NIKKEI-CD} we plot the community structure of the S\&P 500 and Nikkei 225 (daily closing prices from 2001 to 2011) generated using the modified Louvain algorithm. Each community is labelled with the number of stocks and the pie chart represents the relative composition of each community based on the industry sectors of the constituent stocks (color legend in Table I). The links between the communities represent ``residual'' (i.e filtered) anti-correlation relations \cite{MacMahon}. We can see that the binary partition is very similar to the weighted one. For both indexes about $7-8$ percent of the stocks switch community. In the next section we will give a more quantitative measure for the dissimilarities of the different partitions.\\ 

In Figure \ref{FTSE-CD} we observe a more complex result. Again, we plot the community structure of the FTSE100 (daily closing prices from 2001 to 2011) generated using the modified Louvain algorithm. 
Now, the binary representation consistently identifies one more community than the weighted representation (for all the algorithms). Later, we further explore these differences in community structure. 
Most notably, the binary information results in a cluster configuration where the Financials sector (green) was partitioned into two communities, whereas the weighted representation created only one sole Financials community, spreading the rest of the stocks among other clusters.\\

We should note that, purely from a community detection perspective, there is no ``correct'' partition. Each partition is generated from different data and so maximizing modularity should not be expected to yield the same partitions. We are comparing the end results of these processes, and in this setting (this paper) we treat the weighted partition as the ``truth'' since it is using a priori more information to establish the partition. Thus, our aim is merely to examine the degree to which the ``binary'' community structure matches the ``weighted'' community structure, despite having less information for the algorithms to work with. That said, it would be interesting to determine if the binary information can yield different points of view, or added structural information with respect to its weighted counterpart. This would be useful in particular because binary time series are more robust to noise and errors in the data. Such a line of exploration is however beyond the scope of this paper.\\   

In this section we showed that the binary description leads us to a very similar market structure to the weighted description. This results suggest that the information regarding the community partition is mainly encoded in the binary signature of the fluctuations, i.e. just from the knowledge of the direction of movement, one can practically reproduce the ``correct'' structure.
In the next section we will quantify the similarities and deviations of the different partitions for the different algorithms. 
Furthermore, we will explore the evolution in time of the variations between the two representations.

\subsection*{Variation of Information Analysis}
\begin{table}[h!]
	\begin{center}
		\begin{tabular}{|l|l|l|l|l|l|l|l|}\hline
			Index & Method&$\mathbf{Q}$ weighted &$\mathbf{Q}$ binary&Frequent VI&Switching&Minimal VI& Switching\\
			&&&&&stocks (\%)&&stocks (\%) \\ \hline
			 & Potts&0.4035&0.4134&0.3543&8.81\%&0.3198&6.74\%\\
			S\&P & Louvain&0.4070&0.4134&0.3477&8.09\%&0.3192&7.64\%\\
			 & Spectral&0.4006&0.3932&0.6955&61.57\%&0.6955&61.57\%\\  
			 \hline
			 & Potts&0.4551&0.4525&0.3689&6.78\%&0.2598&4.66\%\\
			NIKKEI & Louvain&0.4551&0.4525&0.3711&7.25\%&0.2604&4.15\%\\
			 & Spectral&0.4481&0.4424&0.4521&8.29\%&0.4521&8.29\%\\  
			 \hline
			 & Potts&0.4641&0.4988&0.5031&28.42\%&0.4026&26.14\%\\
			FTSE & Louvain&0.4635&0.4988&0.4995&21.79\%&0.3981&17.95\%\\
			 & Spectral&0.4597&0.4903&0.6919&69.23\%&0.6919&69.23\%\\  
			 \hline
		\end{tabular}
	\end{center}
	\caption{The Variation of Information measured between the binary and weighted partitions, with the maximal modularity $\mathbf{Q}$, for the period 2001-2011. ``Frequent VI'' is the variation of information between the most common partitions (that maximize the modularity), and ``Minimal VI'' is the variation of information between the most similar (that maximize the modularity). The ``Switching stocks'' is the percentage of stocks that moved to different communities.   
	}\label{tab:VarInfo}
\end{table}

Once we obtain the community structure using the different algorithms, our goal is to quantify the dissimilarities (or similarities) between the different partitions (binary and weighted). For this task we apply the Variation of Information (VI) measurement \cite{Meila1,Meila2}. The variation of information is an information-theoretic measure of the distance between two partitions.
The different partitions $\vec{\boldsymbol\sigma^1}$ and $\vec{\boldsymbol\sigma^2}$ represent $N$-dimensional vectors where the $i$-th component $\sigma_i$ denotes the set in which node $i$ is placed by that particular partition.

The variation of information involves the mutual information $I(\vec{\boldsymbol\sigma^1}:\vec{\boldsymbol\sigma^2})$ which is defined as
\begin{equation}
	\label{eq:MutInf}
	I(\vec{\boldsymbol\sigma^1}:\vec{\boldsymbol\sigma^2})=\sum_{i=1}^{N}\sum_{j=1}^{N}p(\sigma^1_i,\sigma^2_j)\textrm{log}\left(\frac{p(\sigma^1_i,\sigma^2_j)}{p(\sigma^1_i)p(\sigma^2_j)}\right),
\end{equation}
where $p(\sigma^1_i,\sigma^2_i)$ is the joint probability distribution, and $p(\sigma^1_i)$ the marginal distribution of $\sigma^1_i$.
The mutual information measures the overlap between the two partitions, however it is not a metric (does not obeys the triangle inequality) nor is it normalized.
Thus, for this study we use the (normalized) variation of information which is defined as  
\begin{equation}
 VI(\vec{\boldsymbol\sigma^1}:\vec{\boldsymbol\sigma^2})= 1-\frac{I(\vec{\boldsymbol\sigma^1}:\vec{\boldsymbol\sigma^2})}{H(\vec{\boldsymbol\sigma^1}:\vec{\boldsymbol\sigma^2})}
\end{equation}
where $H(\vec{\boldsymbol\sigma^1}:\vec{\boldsymbol\sigma^2})$ is the joint entropy and is defined as
\begin{equation}
	\label{eq:MutInf}
	H(\vec{\boldsymbol\sigma^1}:\vec{\boldsymbol\sigma^2})=\sum_{i=1}^{N}\sum_{j=1}^{N}
	p(\sigma^1_i,\sigma^2_j)\textrm{log}\left( p(\sigma^1_i,\sigma^2_j) \right).
\end{equation}

\begin{figure}[h!]
        \centering
        \includegraphics[width=.9\textwidth]{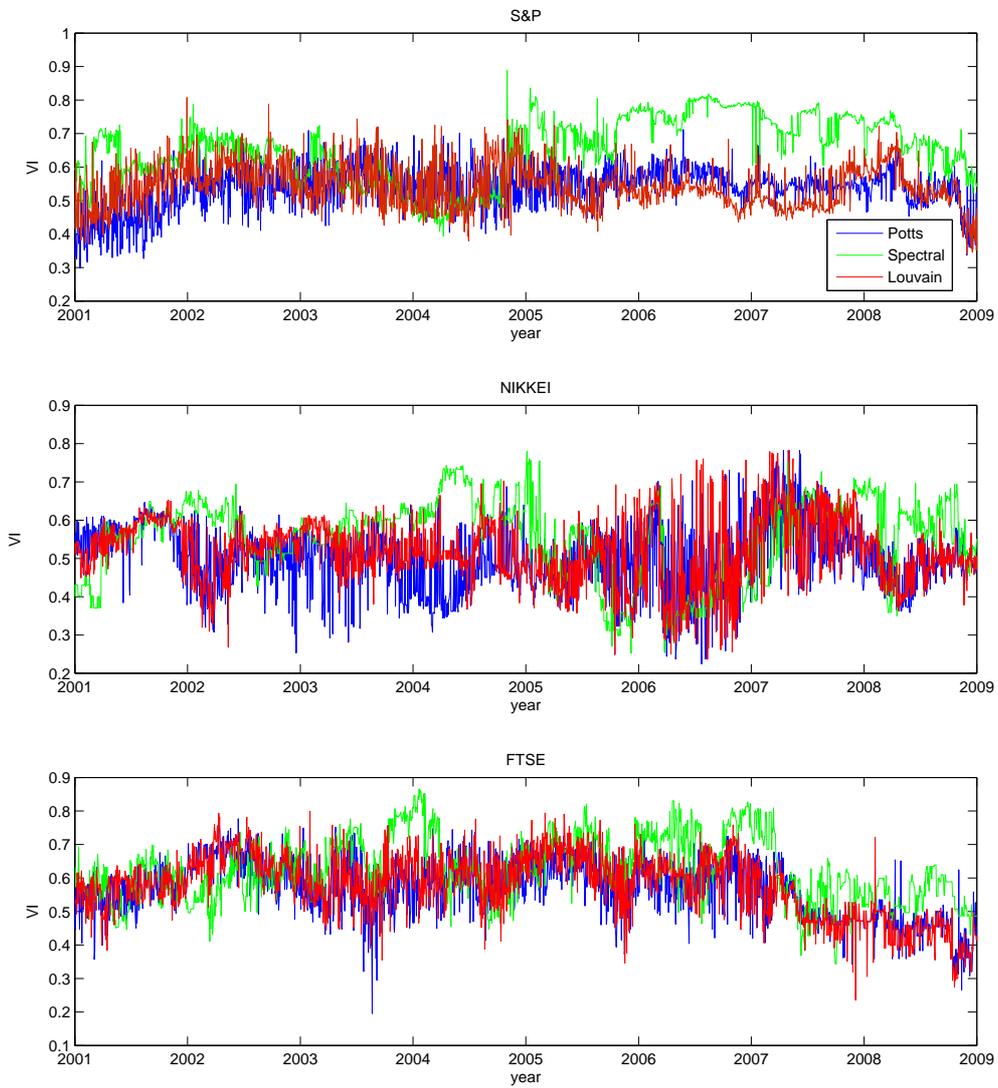}       
        \caption{The variation of information between the binary and weighted partitions for a sliding window of 600 trading days (approximately 28 moths) starting at Q3 2001. The VI is measured between the frequent partitions for the different algorithms: Potts (blue), Louvain(red) and Spectral (green).  }\label{fig:VIslide}
\end{figure}

The variation of information ranges from $0$ to $1$, where $0$ indicates two identical partitions, and $1$ a complete dissimilarity between the partitions.

First, we measure the variation of information between two partition vectors, generated by the weighted and binary time series. 
Respectively, this approach enables us to quantify the difference in group structure (for 2001-2011), and compare the performances of the different algorithms.

In Table \ref{tab:VarInfo} we plot the measured VI between the different partitions, which resulted from the binary and weighted data. These measurements are for the community structures resulting from 10 years (2500 time steps) of data, for each of the three different indices. We run the algorithms 1000 times (for Louvain and Potts, while the spectral is deterministic) and extract the partitions that maximize the modularity.
We measure the VI between two different partitions: the most frequent one, and the one that minimizes the VI (the most similar ones). One can consider this to be the best result (subject to the best partition). Again we should note that all the partitions maximize the modularity, and therefore are optimal. Since the VI is not linear (and not intuitive), we also included a simpler measurement of the percentage of stocks that are not occupying the same community in the different partitions. \\ 

We can observe that both the Potts and Louvain algorithms consistently perform better then the spectral method, i.e. they yield partitions with higher modularity.
We should note that, one can only compare the value of the modularity for the same representation (binary or weighted), while there is no meaning in comparing modularity between the different representations. 
Generally, we can observe that the binary information and the weighted information result in very similar structures, as we showed in the previous section. The exception to this is the FTSE index, where the binary information consistently yields a greater number of communities.
It is interesting to note that the binary information always results in either the same or a greater number of communities over the weighted time series.

Next, we will explore the evolution in time of the VI between the different types of information. We considered a sub-period of 600 time steps (about two and a half years), and apply the same procedure as before. However, here we use a sliding window technique, where in each step we input a new day and ignore the previous information. This results in 1900 time steps (from the original 2500), where each point is the frequent VI calculated from the correlation matrix using the given 600 time step. In Figure \ref{fig:VIslide} we plot the different measurements for the different algorithms.
The Potts and the Louvain methods present a more stable dynamics, while the Spectral method yield higher VI. Furthermore, there is no systematic effects of the financial crisis on the similarity between the binary and the weighted representations.\\

\section*{Discussion}

Over the last few years community detection methods have revealed themselves as useful tools to study the structure of complex systems.
In this context, we have introduced a new approach aiming at analysing structural dependencies, which result form different descriptions (weighted and binary activity) of a complex system.
Our approach enables us to quantify the level of ``structural information'' encoded within the binary projection of weighted time series, and measure variations and similarities between the different partitions.\\ 

The analysis reveals that in financial markets both the binary and weighted information yield very similar structures, which also manifest themselves in similar spectral properties. The algorithms find complex mesoscopic structure of internally correlated clusters, which are residually anti-correlated with each other \cite{MacMahon}.
Moreover, the clusters are populated by stocks from various sectors. Remarkably, we show that the simple knowledge of the direction of increments of each stock can reproduce this complex structure very successfully.\\      

Our findings suggest that the binary signatures of financial time series carry significant structural information. These results are far from trivial, as one might expect that the full knowledge of the amplitudes of price fluctuations is a key component in clustering the markets into correlated groups. However, here we explicitly showed that purely binary information can replicate the main features obtained from complete information. Thus, we conclude that the majority of the market structure induced by the binary dynamics of the stocks. Even when the two representations differ by some extent, the binary description provides very sensible information (as exemplified by the Financials sector in the FTSE).
In any case, binary projections are much more robust to noise then the original fluctuations, by definition.

\section*{Methods}

\subsection*{Community detection}
\label{CD}

We adopt modularity-based community detection methods, which are adapted to correlation matrices.
Respectively, this restrict us to undirected networks, as a result of sharing by definition the same symmetry property as correlation matrices.
Let us consider a network with $N$ nodes, one can introduce a number of partitions of the $N$ nodes into non-overlapping sets. 
The different partitions will be represented by an $N$-dimensional vector $\vec{\sigma}$  where
the $i$-th component $\sigma_i$ denotes the set in which node $i$ is placed by that particular partition. 
Now, we introduce the modularity measure $Q(\vec{\boldsymbol\sigma})$ which indicates the quality of a specific choice of partition $\vec{\sigma}$ measured by high degree of inter community connectivity and a low degree of intra community connectivity. The modularity optimization algorithms look for the specific partition that maximizes the value of $Q(\vec{\boldsymbol\sigma})$, the objective function.
It is defined as
\begin{equation}Q(\vec{\boldsymbol\sigma})=\frac{1}{A_{tot}} \sum_{i,j} \left[ A_{ij}-\langle A_{ij} \rangle \right]\delta(\sigma_i, \sigma_j)
\end{equation}
where $\delta (\sigma_i , \sigma_j )$ is a delta function ensuring that only when $\sigma_i= \sigma_j$ (nodes
within the same community) does it contribute to the sum, and $A_{ij}$ is the adjacency matrix that indicates whether a link exists between the nodes, $A_{ij}=1$ or not, $A_{ij}=0$ (in the binary representation). The pre-factor $A_{tot}$ serves to normalize the value of $Q(\vec{\boldsymbol\sigma})$
between $-1$ and $1$, where $A_{tot}\equiv \sum_{i,j} A_{ij}=2L$ is twice the number of total links in the network. 
The term $\langle  A_{ij} \rangle$ is vital to the outcome of the community detection process. 
It represents the expectation of whether a link exists or not, according to the specific null model that you consider.
So far the majority of the methods use null models (hypotheses), which are suited only for networks.
For example the configuration model, that preserves the degree sequence (or strength sequence) of the network. 
It has been shown that such null models can introduce biases when applied to correlation matrices \cite{MacMahon}.
Instead, a recent method proposed a redefinition of the modularity, which takes into account the existence of known spectral properties in correlation matrices (see section \ref{SA}).\\ 

Now, instead of the previous adjacency matrix $A_{ij}$, we input the empirical correlation matrix $C_{ij}$.
The first is the global mode $C^{(m)}$(market mode in a financial setting), which represents the common movement of the market. In other words, in order to clearly differentiate between the mesoscopic groups, one must subtract out the main drift of the market. 
The second is the random bulk $C^{(r)}$
\begin{equation}
 C^{(r)}=\sum_{i:\lambda_i \leq \lambda+} \lambda_i |v_i\rangle \langle v_i|
\end{equation}
which corresponds to random correlation between the different time series. 
In order to filter this noise, one must use random matrix theory (RMT) \cite{Utsugi},to identify the random properties of empirical correlation matrices. 
The method define the modularity as 
\begin{equation}
	\label{eq:Modularity}
	 Q(\vec{\boldsymbol\sigma})=\frac{1}{C_{norm}}\sum_{i,j}(C_{ij}-C_{ij}^{(r)}-C_{ij}^{(m)})\delta(\sigma_i,\sigma_j)=\frac{1}{C_{norm}}\sum_{i,j}C_{ij}^{(g)}\delta(\sigma_i,\sigma_j)
\end{equation} 
constituted from the eigenvalues $\{\lambda_i\}$ less than or equal to $\lambda+$ (usually, the eigenvalues
smaller than $\lambda-$ are included as well) and their corresponding eigenvectors ${v_i}$.
The new method modified three popular community detection algorithms, customizing where necessary to be effective with correlation matrices \cite{MacMahon}. The three algorithms we use in this paper are known as the Potts (or spin glass) method \cite{Potts1,Potts2}, the Louvain method \cite{Louvain} and the spectral method \cite{Spectral}

\section*{Acknowledgments}

AA and DG acknowledge support from the Dutch Econophysics Foundation (Stichting Econophysics, Leiden, the Netherlands) with funds from beneficiaries of Duyfken Trading Knowledge BV, Amsterdam, the Netherlands. 
This work was also supported by the EU project MULTIPLEX (contract 317532) and the Netherlands Organization for Scientific Research (NWO/OCW).

\section*{Author Contributions}

A.A. and F.B. analyzed the data and prepared the figures. A.A. wrote the paper.
M.M. wrote the code and created figures 3-5.  
D.G. planned the research and supervised the project. All authors reviewed the manuscript.

\section*{Competing financial interests}

The authors declare no competing financial interests.

\end{document}